\documentclass[11pt, letterpaper, twoside]{article}
\usepackage[utf8]{inputenc}
\usepackage[T1]{fontenc}
\usepackage[babel=true]{microtype}
\usepackage{siunitx, dsfont}
\usepackage[main=english, spanish]{babel}
\usepackage{subcaption, booktabs, sectsty}
\usepackage{fancyhdr}
\usepackage[affil-it]{authblk}
\usepackage[colorlinks=true, citecolor=blue, pagebackref=true, linktocpage]{hyperref}
\usepackage{tikzsymbols, cite, url, csquotes}
\usepackage{amsmath, amsfonts, amssymb, amsthm, amsbsy, mathtools, mathrsfs, xfrac, physics}
\usepackage{array, multirow, multicol}
\usepackage{graphicx, float, comment, color}
\usepackage[labelfont=bf, font=small]{caption}
\usepackage[left=2.5cm, right=2.5cm, top=2.5cm, bottom=2.5cm]{geometry}
\colorlet{linkequation}{blue}
\hypersetup{
    colorlinks,
    citecolor=blue,
    filecolor=black,
    linkcolor=blue,
    urlcolor=blue
}

\newcommand*{\SavedEqref}{}
\let\SavedEqref\eqref
\renewcommand*{\eqref}[1]{%
  \begingroup
    \hypersetup{
      linkcolor=linkequation,
      linkbordercolor=linkequation,
    }%
    \SavedEqref{#1}%
  \endgroup
}

\def\al{\alpha}
\def\b{\beta}
\def\th{\theta}  
\def\lam{\lambda}

\def\D{\Delta}

\def\chic#1{{\scriptscriptstyle #1}}

\def\be{\begin{equation}}
\def\ee{\end{equation}}
\def\bea{\begin{eqnarray}}
\def\eea{\end{eqnarray}}
\def\ba{\begin{array}}
\def\ea{\end{array}}

\def\hs{\hspace}

\def\hsp{\hspace{.03cm}}
\def\hsn{\hspace{-0.03cm}}
\def\ben{\begin{enumerate}}
\def\een{\end{enumerate}}
\def\bei{\begin{itemize}}
\def\eni{\end{itemize}}

\def\mc{\mathcal}
\def\ms{\mathscr}
\def\mtt{\mathtt}

\def\lan{\langle}
\def\ran{\rangle}
\def\dag{\dagger}
\def\bco{}

\begin{document}

\title{\textbf{Interplay of non-standard interactions and Earth's composition in atmospheric neutrino oscillations}}

\author[1,2]{\normalsize Juan Carlos D'Olivo%
	\thanks{email: \texttt{dolivo@nucleares.unam.mx}}} 

\author[1]{\normalsize Jos\'e Arnulfo Herrera Lara%
	\thanks{email: \texttt{fis\_pp@ciencias.unam.mx}}} 

\author[2]{\normalsize Ismael Romero%
	\thanks{email: \texttt{ismaelromero\_@hotmail.com}}}

\author[2]{\normalsize Matias Reynoso%
	\thanks{email: \texttt{matiasreynoso@gmail.com}}}

\author[2]{\normalsize Oscar A. Sampayo%
	\thanks{email: \texttt{sampayo@mdp.edu.ar}}}

\affil[1]{\normalsize Instituto de Ciencias Nucleares, Universidad Nacional Aut\'onoma de M\'exico, 
Circuito Exterior, Ciudad Universitaria, 04510 CDMX, Mexico.}

\affil[2]{\normalsize Instituto de F\'isica de Mar del Plata (IFIMAR)\\
CONICET, UNMDP\\ Departamento de F\'isica,
Universidad Nacional de Mar del Plata \\
Funes 3350, (7600) Mar del Plata, Argentina}

\date{}

\maketitle

\begin{abstract}
Many geophysical and geochemical phenomena in the Earth's interior are related to physical and chemical processes in the outer core and the core-mantle boundary, which are directly linked to the isotopic composition. Determining the composition using standard geophysical methods has been challenging. Atmospheric neutrino oscillations, influenced by their weak interactions with terrestrial matter, offer a new way to gather valuable information about the Earth's internal structure and, in particular, to constrain the composition of the core. If neutrinos had as yet unknown non-standard interactions (NSI), this could affect their propagation in matter and consequently impact studies of Earth's composition using neutrino oscillation tomography. This study focuses on scalar-mediated NSI and their potential impact on atmospheric neutrino oscillations, which could obscure information about the hydrogen content in the outer core. In turn, compositional uncertainties could affect the characterization of NSI parameters. The analysis is based on a Monte-Carlo simulation of the energy distribution and azimuthal angles of neutrino-generated $\mu$ events. Using a model of the Earth consisting of 55 concentric shells with constant densities determined from the PREM, we evaluate the effect on the number of events due to changes in the outer core composition (Z/A)$_{oc}$ and the NSI strength parameter $\epsilon$. To examine the detection capability to observe such variations, we consider regions in the plane of (Z/A)$_{oc}$ and $\epsilon$ where the statistical significance of the discrepancies between the modified Earth model and the reference model is less than $1\sigma$.
\end{abstract}

\section{Introduction}
\label{intro}
Seismic studies indicate that the Earth has a spherical structure of concentric shells whose density steadily increases as one descends toward the center. From the analysis of iron meteorites and the observation of the planet's moment of inertia, it is known that the Earth's core is composed primarily of an iron-nickel alloy, with Ni/Fe\,$\sim \num{0.06}$. Two well-differentiated regions have been identified: the inner and outer core, distinguished by an abrupt density difference of \SI{4.5}{\percent} at a depth of approximately \SI{5100}{\kilo\meter} \cite{Clement:1997}. Seismic velocity data indicate that the outer core does not transmit S-waves \cite{Jeanloz:1990, Jacobs:1992, Tkalcic:2008}. This observation, along with the lower density, is interpreted to mean that the outer core is in a liquid state. The inner core transmits S-waves at very low velocities, suggesting it is a solid near the melting point or partly molten. The density deficit in the outer core is too large to be explained simply by a solid-liquid phase transition, and it requires about 5-10 wt\% (weight percent) of lower atomic weight elements to reduce its density and melting point \cite{Birch:1964, Poirier:1994}. Reasonable estimates of the abundance and distribution of ``light''  elements in the core are essential to understanding the formation and evolution of the Earth, as well as how the core and mantle interact in the region around the core-mantle boundary (CMB) \cite{Litasov:2016,Hiroshe:2021}. 

Since F. Birch \cite{Birch:1964} suggested that pure Fe or Fe-Ni alloy alone is too dense, several light elements have been proposed as candidates existing in the Earth's core, including O, Si, S, C, and H; ascertaining their varieties and concentrations has been a longstanding challenge. While the density distribution can be obtained from seismological 
remote sensing, the compositional structure of the Earth \cite{ Allegre1995, McDonough:1995} has been much more challenging to determine. Thus, the compositions of the lower mantle and core remain uncertain despite significant advances in recent years. The estimates compare the density and sound velocity data from seismological observations with those from laboratory experiments and theoretical calculations \cite{Zhang2016}. It is difficult to conduct experiments with samples of liquid iron alloy under the extreme temperature and pressure conditions of the core \cite{Morard:2013} and, as a consequence, the available accurate experimental data are insufficient to be compared with seismological data directly. In recent work, the chemical composition of liquid iron alloys at outer-core pressure and temperature conditions has been constrained by comparing first-principle calculations of the density and bulk sound velocity with the profiles of the Preliminary Reference Earth Model (PREM) \cite{umemoto:2020}. The primary light component was found to be hydrogen when the inner-core boundary temperature $T_{\scriptscriptstyle \mathrm{ICB}}$ is not high (\SI{4800}{\kelvin} to \SI{5400}{\kelvin}) or oxygen if $T_{\scriptscriptstyle \mathrm{ICB}} = \SI{6000}{\kelvin}$. Hence, as the authors concluded, it remains challenging to specify the outer-core chemical composition by comparing only theoretical calculations and seismological observations of density and speed of sound.

Atmospheric neutrinos are a promising and viable experimental technique for obtaining complementary and independent information about the Earth's interior. They are produced isotropically in the upper atmosphere as decay products in hadronic showers generated by cosmic-ray (usually protons) collisions with air nuclei. This provides a continuous source of neutrinos with travel distances from 15 to 12757 \unit{\kilo\meter} to the detector and a wide range of energies from MeV to tens of TeV, making them well-suited for tomographic studies of the planet \cite{Winter:2006vg}.
In neutrino absorption tomography, the density profile can be reconstructed from he attenuation of the flux of very high-energy neutrinos ($\gtrsim 10$ TeV), which can undergo inelastic scattering and be absorbed as they pass through the Earth \cite{Volkova:1974xa, Wilson:1983an, Ralston:1999fz, Jain:1999kp, GonzalezGarcia:2007gg, Reynoso:2004dt, Romero:2011zzb, Donini:2018tsg}. The feasibility of this approach has been demonstrated using real data collected by the IceCube telescope \cite{Donini:2018tsg}. Another option is oscillation tomography, which exploits the effect of matter on the flavor oscillations of lower energy neutrinos (from MeV to GeV), particularly those with energies in the interval from 1 to 10 GeV, where these effects are most significant \cite{Nicolaidis:1987fe, Nicolaidis:1990jm, Borriello:2009ad, Winter:2015zwx, Rott:2015kwa, Bourret:2017tkw, VanElewyck:2017dga, DOlivo:2020ssf, Kelly:2021jfs, {Denton:2021rgt}, DOlivoSaez:2022vdl, Upadhyay:2023kzf, Maderer:2023toi}.

After decades of experimental observations and theoretical developments, the $3\nu$-oscillation paradigm has been firmly established as the mechanism underlying neutrino flavor transitions. In a medium, due to coherent forward scattering with background particles, the probabilities of the flavor transformations depend not only on the neutrino energy and the traveled distance $z$, but also on the electron number density 
$n_e(z)$ along the neutrino path \cite{Wolfenstein:1977ue, Mikheyev:1985zog}: 
\be
\label{eq:elecdensity}
n_e (z) = \left(\dfrac{Z}{A}\right)\!\!(z) \dfrac{{\rho}(z)}{m_{\rm u}},
\ee
where $\rho$ is the matter density, $m_{\rm u} = \SI{931.494}{\mega\electronvolt}$ is the atomic mass unit, and $Z/A$ is the average ratio between the atomic number $Z$ and the (relative) atomic mass $A$. For a compound $(Z/A)(z) = \sum_{\lam} {\mtt r}_{\!\lam}\hsn(z)(Z/A)_{\lam}$, where the summation runs over all the elements contributing a fraction $\mtt{r}_{\!\lam}$ to the total mass. Therefore, it might be possible to constrain the permissible values of the density and composition of the deep parts of the Earth by studying how changes in these quantities affect the number of events produced in a detector by neutrinos after they pass through the Earth. Except for hydrogen ($(Z/A)_H = 1$), all other light elements have nearly equal protons and neutrons and hence $Z/A \cong \num{0.5}$ for them.  Consequently, a relatively small amount of hydrogen would cause a noticeable change in $n_e$, and neutrino oscillations could provide valuable information about the presence of this element in the core.

Out of the six neutrino oscillation parameters, $\D m^2_{21}$, $|\D m^2_{31}|$, with $\D m^2_{ij} = m^2_i - m^2_j$, and the mixing angles $\th_{12}$, $\th_{13}$, $\th_{23}$ of the matrix PMNS are known with ever-increasing accuracy \cite{ParticleDataGroup:2024cfk}. The sign of $\D m^2_{31}$ characterizes the normal ordering (NO), with $m_3 > m_1$, and the inverted ordering (IO), with $m_3 < m_1$.  Several experiments are already underway or in preparation to determine the ordering of neutrino masses and to address the other two outstanding questions: the value of the CP-violating phase $\delta$ and the octant of $\th_{23}$ \cite{jamieson:2022}.  At the same time, these experiments will be sensitive to sub-leading effects associated with unknown interactions of neutrinos with ordinary matter parameterized at low energies by effective four-fermion couplings, which are generally referred to as non-standard neutrino interactions (NSI) (for reviews, see \cite{Ohlsson2013, Miranda:2015, farzan:2018}). They can affect neutrino oscillation signals through charged current (CC) processes, which impact neutrino production and detection, or through neutral current (NC) processes, which influence neutrino propagation in matter. Such extensions of the Standard Model (SM) have attracted the community's attention due to the impact the additional interactions can have on getting the values of the oscillation parameters \cite{Capozzi:2019iqn, Esteban:2019}.  Also, flavor-changing NC-NSI have been considered as a possible solution to the tension between the latest measurements of the CP phase from the long-baseline accelerator experiments NOvA and T2K \cite{PhysRevLett.126.051801, Chatterjee:2020kkm}.

In this work, we consider the application of atmospheric neutrino oscillations to examine Earth's internal composition, including potential NSI effects. Specifically, we seek to determine to what extent NC-NSI effects on neutrino oscillations might affect the determination of the H content in the outer core and, vice versa, how compositional uncertainties might mask effects associated with non-standard neutrino interactions. To do so, we analyze the ability of a large detector such as ORCA to resolve deviations in the number of $\mu$-like events due to changes in the flavor content of the atmospheric neutrino flux as a consequence of the modifications that a combination of the aforementioned effects can have on neutrino oscillations in matter.

The paper is organized as follows. In section \ref{sec:structure_osc}, we briefly review matter neutrino oscillations and describe the calculation of the transition probabilities, including NSI effects, for atmospheric neutrinos passing through the Earth, modeled as a sphere divided into 55 concentric shells integrated into five main layers. In section \ref{sec:simu}, we perform a Monte Carlo simulation of the number of $\mu$-like events in a detector like ORCA. The results and final comments are presented in section \ref{sec:results}, where we use our calculations to constrain the intensity of the non-standard interactions and the hydrogen content in the outer core.

\section{Oscillations of atmospheric neutrinos}
\label{sec:structure_osc}

Neutrino dispersion relations are modified in a medium due to their coherent forward scattering with the background particles: in normal matter, electrons ($e$), up quarks ($u$), and down quarks ($d$).The scattering of $\nu_e$ with electrons differs from that of  $\nu_{\mu,\tau}$ and, after subtracting an unobservable overall phase, this introduces an additional term in the Hamiltonian $H(z)$ that governs the evolution of the flavor amplitudes along the neutrino trajectory in matter. As a result, the pattern of neutrino oscillations can change in a non-trivial manner compared to vacuum oscillations, and resonance enhancement effects can appear under appropriate conditions. 
\begin{figure*}[!ht]
\centering
\includegraphics[width=0.30\textwidth]{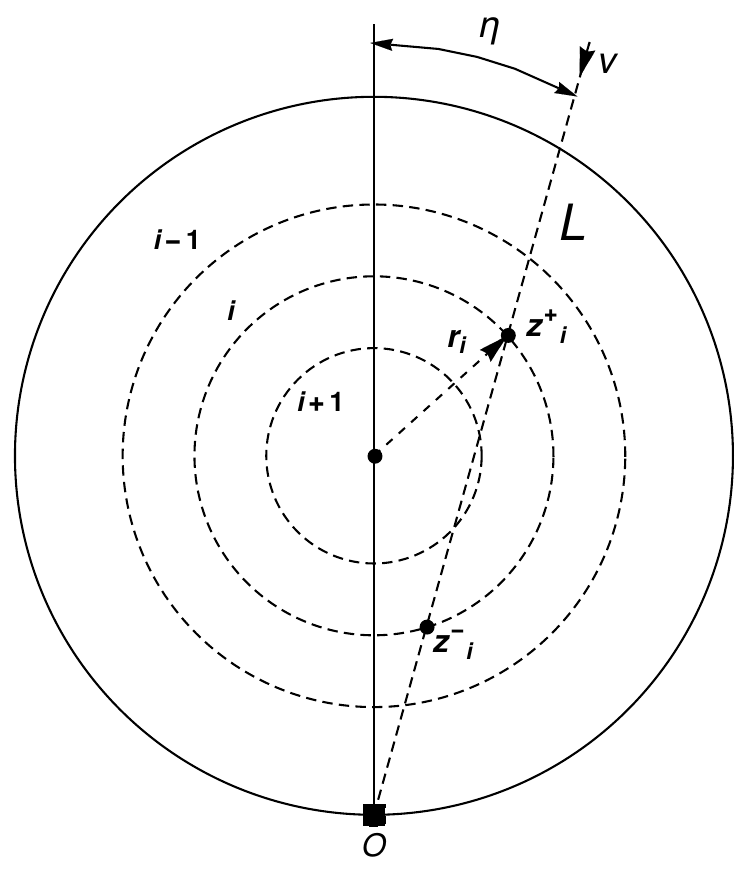}	
\label{}	
\hfill
\caption{Neutrino path through the Earth.}
\label{fig:earth}
\end{figure*}

Nonstandard neutral-current (NC) interactions may contribute to forward scattering processes and affect neutrino flavor transformations in the medium.  At low energies, this can be conveniently studied within the frame of effective field theories parametrized in terms of four-fermion operators \cite{Wolfenstein:1977ue, Mikheyev:1985zog}. 
The NC-NSI effective Lagrangian at the quark level reads,
\begin{equation}
\label{eq:NSI_Lagrangian}
\ms{L}_{\chic {NSI}}=-2\sqrt{2}\,G_{F} \!\!\sum\limits_{f,\alpha, \beta}\epsilon_{\alpha\beta}^{f}(\bar{\nu}_\alpha\gamma^{\mu}P_L\nu_\beta)
(\bar{f}\gamma^{\mu}f),
\end{equation}
where $G_{F}$ is the Fermi constant, $P_{L} = \tfrac{1}{2}(1 - \gamma_5)$, and the sum runs over the neutrino flavors ($\alpha, \beta = e, \mu, \tau$) and the charged fermions ($f = e, u,d$). We have assumed that the neutrino flavor structure of the interactions is independent of the charged fermion type \cite{Esteban:2018}.  The dimensionless coefficients $\epsilon_{\al\beta}^{f}$ quantify the strength of the NSI relative to the SM weak interaction: $\epsilon_{\al\beta}^{f}\sim O(G_X/G_F)$. Hermiticity of $\ms{L}_{\chic {NSI}}$ implies these coefficients satisfy $\epsilon_{\beta\alpha}^{f}=\epsilon_{\alpha\beta}^{f*}$ and hence, those with $\al = \beta$ are real.  

For neutrinos propagating in ordinary matter, the contributions of the NC-NSI to the effective potential in the flavor evolution equation are:
\be
\label{eq:potentialnsi}
\sqrt 2 G_F\!\sum_{f} n_f(z)\epsilon^f_{\al\beta} = {\mc V}(z)\!\left[\epsilon_{\al \beta} + \left(\frac{n_n(z)}{n_p(z)} - 1\right)\epsilon^n_{\al \beta}\right],
\ee
where $\epsilon_{\al \beta} \equiv \epsilon^e_{\al \beta} + \epsilon^p_{\al \beta} + \epsilon^n_{\al \beta}$, $n_{p, n}(z)$ are the number densities of protons and neutrons, and 
${\mc V} (z) = \sqrt 2 G_F n_e(z)$, with $n_e(z)$ the electron number density given in Eq. \eqref{eq:elecdensity}. When writing the right-hand side of the previous equation, we used the relations $n_u = 2n_p + n_n$ and $n_d = n_p + 2 n_n$, and made the identifications $\epsilon^p_{\al\beta}\ = 2 \epsilon^u_{\al\beta} + \epsilon^d_{\al\beta}$ and $\epsilon^n_{\al\beta} = 2 \epsilon^d_{\al\beta} + \epsilon^u_{\al\beta}$. We also considered that, due to the neutrality of matter, $n_e(z) = n_p(z)$.  

For Earth, the values of $n_n/n_p$ given by the PREM are \num{1.012} in the mantle and \num{1.137} in the core. Therefore, assuming that the coefficients $\epsilon_{\al \beta}$ and $\epsilon^n_{\al \beta}$ have similar magnitudes, the second term in Eq. \eqref{eq:potentialnsi} is about ten times smaller than the first and will be ignored to keep our analysis as simple as possible. Thus, the Hamiltonian governing the evolution of the flavor amplitudes of atmospheric neutrinos traveling within terrestrial matter, in the basis of mass eigenstates, is given by:                                                                                                                                                                                                                                                                                                              
\be
\label{hamiltonian}  
\begin{aligned}
H(z) &= \begin{pmatrix}            
            0   &          0         &      0   \\
            0   &    \hs{0.2cm}{\D}_{21}  &      0   \\            
            0   &          0         &      {\D}_{31}
\end{pmatrix}   \\        
&\phantom{=}+
{\mc V}(z) U^{\dag}\hs{-0.12cm}\begin{pmatrix}
            1+\epsilon_{ee}-\epsilon_{\mu\mu}    & \;\;\epsilon_{e\mu} &  \epsilon_{e\tau}   \\
            \epsilon_{\mu e}   &      0  &    \epsilon_{\mu\tau}   \\
            \epsilon_{\tau e}   &      \epsilon_{\tau\mu}   & \;\;\epsilon_{\tau\tau}-\epsilon_{\mu\mu} 
\end{pmatrix}\hs{-0.08cm}U\,,
\end{aligned}
\ee
where $U$ is the PMNS mixing matrix and the ``one'' in the $ee$ entry of the effective potential matrix accounts for the standard charged-current contribution. In writing $H(z)$, the usual non-relativistic approximation $E_i- E_j\cong \D_{ij}$ was used, and the $\mu\mu$ element of the potential energy matrix was subtracted by factoring a global phase proportional to $\epsilon_{\mu\mu}$.
\begin{table}[!ht]
\centering
\begin{tabular}{@{}ccc@{}}
\toprule
\textbf{Parameter}      & \textbf{Normal Ordering}       & \textbf{Inverted Ordering}    \\ \midrule
$\Delta m^{2}_{21}$ [\unit{\square\electronvolt}]         & \num{7.42e-5}   & \num{7.42e-5}   \\[1.0pt]
$\Delta m^{2}_{31}$ [\unit{\square\electronvolt}]      & \num{+2.533e-3} & \num{-2.437e-3} \\[1.0pt]
$\sin^{2}{\theta_{12}}$                                   & \num{0.309}     & \num{0.308}     \\[1.0pt]
$\sin^{2}{\theta_{13}}$                                   & \num{0.0223}    & \num{0.0232}    \\[1.0pt]
$\sin^{2}{\theta_{23}}$                                   & \num{0.561}     & \num{0.564}     \\[1.0pt]
$\delta/\pi$                                              & \num{1.19}      & \num{1.54}  \\ \bottomrule
\end{tabular}
\caption{Three-neutrino oscillation parameters obtained by averaging the best-fit values of three recent global fits of the current neutrino oscillation data \cite{Capozzi:2020,Esteban:2020cvm,deSalas:2020pgw}.}
\label{tab:constants}
\end{table}
The probabilities  $P_{{\nu_\al} \to {\nu_\b}}(z)$ that a neutrino produced as a $\nu_{\al}$ is found in a flavor $\nu_{\b}$ at a distance $z$ from the source can be written as:
\be
\label{oscprob}
P_{{\nu_\al} \to {\nu_\b}}(z) = |\,{\mc U}_{\b \al}(z)|^2\,,
\ee
where 
$
\label{matrixelem}
{\mc U}_{\b \al}(z) = \lan \nu_{\b}|\,\hat {\mc U}(z)|\nu_{\al}\ran
$
are the elements of the $3 \times 3$ matrix ${\mc U}_f(z)$ representing the evolution operator 
in the flavor basis. This matrix is related to that in the mass eigenstates basis, ${\mc U}_m(z)$, 
by the unitary transformation 
\be
\mc U_f(z) = U \mc{U}_{m}(z) U^{\dagger}, 
\ee
where $\mc{U}_{m}$ has to be determined by solving the equation
\be
\label{eq:Scheq}
i \hs{.02cm}\dv{z} \mc{U}_{m}(z) = H(z) \, \mc{U}_{m}(z),
 \hs{0.7cm} \mc{U}_{m}(0) = I, 
\ee
with $H(z)$ given by the matrix in Eq. \eqref{hamiltonian}. 
\begin{figure*}[!htbp]
  \centering
  \begin{subfigure}[b]{0.48\textwidth}
    \centering
    \includegraphics[width=\textwidth]{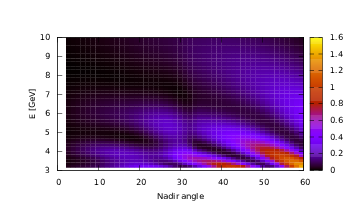}
    \caption{True events, normal mass ordering}
    \label{fig:eve_ver_NO}
  \end{subfigure}
  \hfill 
  \begin{subfigure}[b]{0.48\textwidth}
    \centering
    \includegraphics[width=\textwidth]{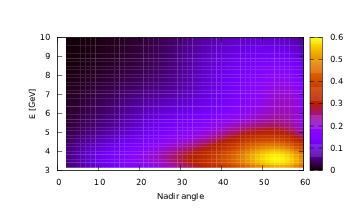}
    \caption{Observed events, normal mass ordering}
    \label{fig:eve_obs_NO}
  \end{subfigure}
    \vspace{0.5cm} 
  \begin{subfigure}[b]{0.48\textwidth}
    \centering
    \includegraphics[width=\textwidth]{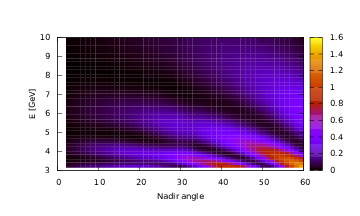}
    \caption{True events, inverted mass ordering}
    \label{fig:eve_ver_IO}
  \end{subfigure}
  \hfill 
  \begin{subfigure}[b]{0.48\textwidth}
    \centering
    \includegraphics[width=\textwidth]{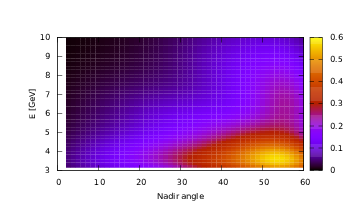}
    \caption{Observed events, inverted mass ordering}
    \label{fig:eve_obs_IO}
  \end{subfigure}
  \caption{Probability distribution for true and observed events.}
    \label{fig:true_obs_NO_IO}
\end{figure*}

In the present study, we employ a spherically symmetric one-dimensional model of our home planet consisting of 55 concentric shells, each with a constant density equal to the average value of the PREM densities in the shell. The set of shells is divided into five main layers bounded by concentric spheres of different radii: inner core ($IC$), outer core ($OC$), lower mantle ($M_1$), upper mantle ($M_2$), and crust ($C$). The compositions and radii of these layers are given in Table 2. We use a single value of $Z/A$ for the entire mantle, as well as for the inner and outer core. The values of the radii are those of the PREM. Following the approach of Refs. \cite{DOlivo:2020ssf, DOlivoSaez:2022vdl}, we require that the sum of the masses and the sum of the inertial moments of the main spherical layers be equal to the total mass of the Earth 
$
M_{_{\hsn\oplus}} = 5.9724 \times 10^{27} {\rm g}
$
and its mean moment of inertia about the polar axis 
$
I_{_{\hsn\oplus}} = 8.025 \times 10^{44} {\rm g} \hs{0.03 cm}{\rm cm}^2.
$ 
We will refer to this model as the {\it standard Earth} and compare it to others with different compositions in the outer core 
to appreciate the effects that a significant amount of hydrogen in this layer can have.
\begin{table}[h]
\centering
\begin{tabular}{@{}cccc@{}}
\toprule
\textbf{Layer} & \textbf{No. of Shells} & \textbf{$R_{\mathrm{inf}}$ - $R_{\mathrm{sup}}$ (\unit{\kilo\meter})} & \textbf{Z/A} \\ \midrule
Inner Core     & 7                      & 0 - 1221.5                       & \num{0.4691}       \\[1.0pt]
Outer Core     & 13                     & 1221.5 - 3480                    & \num{0.4691}       \\[1.0pt]
Lower Mantle   & 6                      & 3480 - 4200                      & \num{0.4954}       \\[1.0pt]
Upper Mantle   & 23                     & 4200 - 6346                      & \num{0.4954}       \\[1.0pt]
Crust          & 6                      & 6346 - 6371                      & \num{0.4956}       \\ \bottomrule
\end{tabular}
\caption{Number of shells, radius, and composition of the main layers of the Earth.}
\label{tab:compo}
\end{table}

The complete evolution operator can be expressed as the product of the evolution operators of the consecutive shells through which the neutrinos pass on their way to the detector:
\begin{eqnarray}
\label{fullevolop}
\mc U_m(L) = \prod\limits_{j=1}^{2 j_m-1} \mc U_m^j(L_j)\,,
\end{eqnarray} 
with $L = \sum_j L_j$, where $L_j$ is the path length in shell $j$, which
depends on the nadir angle $\eta$. From Fig. \ref{fig:earth}, it can be seen that
$L_j=\vert \tilde{z}_{j+1} - \tilde{z}_j \vert$,
where
\begin{equation}
 \tilde{z}_j = \begin{cases}
z^+_j\,,  & 1 \le j \le j_m\,,\\
z^-_{2j_m +1 -j}\,, & j_m + 1 \le j \le 2j_m\,,
\end{cases}
\end{equation}
\begin{equation}
z^{\pm}_ j = R_{_{\hsn\oplus}} \cos\eta\pm \sqrt{r_j^2-(R_{_{\hsn\oplus}} \sin\eta)^2}\,.
\end{equation}
The effective potential in each shell takes the fixed value $ {\mc V}_{\!j} = \sqrt{2}\hsp G_{\hs{-.03cm}F} n^{j}_e,$ where  $n^{j}_e$ denotes the constant number density of electrons in the $j$ shell. The corresponding evolution operator is 
\be
\mc U_m^j(L_j) = \exp(-i H_{\!j} L_j)\,.
\ee
Since the Hamiltonians for the different layers generally do not commute with each other, the exponential factors in Eq. \eqref{fullevolop} must go in the prescribed order. Each of these factors can be evaluated by applying the Cayley-Hamilton theorem, which allows us to convert the infinite series into a polynomial:
\be
\mc U^{j}_m(L_{j})  = a^j_0 I+a^j_1 {H}_{\!j} + a^j_2 {H_{\!j}}^2 \,,
\ee
where the coefficients are functions of the eigenvalues of ${H}_{\!j}$ 
\cite{Ohlsson:1999xb, DOlivo:2020ssf}.

%
The relevant quantities for us are the oscillation probabilities for ${\nu_\mu} (\bar{\nu}_\mu) \to {\nu_{\mu.\tau}} (\bar{\nu}_{\mu,\tau})$ and ${\nu_e} (\bar{\nu}_e) \to {\nu_{\mu,\tau}} (\bar{\nu}_{\mu,\tau})$, which are needed to compute the $\mu$-like events produced in a detector by ``upward'' atmospheric neutrinos after traveling a distance $L$ through the Earth. Due to the dependence of these probabilities on the effective potential energy ${\mc V}(z)$, they are sensitive to changes in the density and composition of the traversed internal regions of the Earth.  In what follows, we analyze the feasibility of applying this effect to obtain meaningful information about the hydrogen content in the outer core using atmospheric neutrino oscillations in the presence of NSI.
%
\section{Muon events in a Cherenkov detector}
\label{sec:simu}
Let us consider a generic detector with a water or ice mass containing $n_{\chic N}$ nucleons as a target, which can efficiently detect Cherenkov radiation emitted along the trajectories of the  $\mu^{\pm}$ produced by the interactions of $\nu _ {\mu}$ and $\bar\nu_{\mu} $ with nucleons in or near the instrumented volume. It can also detect neutrino showers due to the $e$ or $\tau$ produced by the associated neutrinos \cite{Stavropoulos:2021hir,Kalaczynski:2021ytv,Bourret:2017tkw}. To generate a data set that resembles real observations, with the inevitable statistical fluctuations, we performed a Monte Carlo simulation of the neutrino events, employing the same procedure implemented in \cite{DOlivo:2020ssf}, which we summarize below.

Our observable is the number ${\mathcal N}_{\mu}$ of $\mu$-like events within given angular and energy intervals:
\begin{equation}
\label{eq:numeroeventos}
\begin{aligned}
\mathcal{N}_{\mu} &= n_{N\scriptscriptstyle} T \!\!\int_{E_{\text{min}}}^{E_{\text{max}}} \!\int_{\eta_{\text{min}}}^{\eta_{\text{max}}} \dd{E} \dd{\Omega} \times \\ 
        &\hs{-0.15 cm}\left[\sigma^{cc}_{\nu_{\mu}}(E) \left(P_{\nu_{\mu}\rightarrow \nu_{\mu}} {\dfrac{d\Phi}{dE}}^{\!\!\nu_{\mu}} + P_{\nu_{e}\rightarrow \nu_{\mu}}{\dfrac{d\Phi}{dE}}^{\!\!\nu_{e}}\right) + \right. \\
	& \left. \sigma^{cc}_{\bar\nu_{\mu}}(E)\left(P_{\bar \nu_{\mu}\rightarrow \bar \nu_{\mu}} {\dfrac{d\Phi}{dE}}^{\!\!\bar \nu_{\mu}} + P_{\bar \nu_{e}\rightarrow \bar \nu_{\mu}} {\dfrac{d\Phi}{dE}}^{\!\!\bar \nu_{e}} \right) + \right. \\
	& \left. \sigma^{cc}_{\nu_{\tau}}(E) Br_{\tau \rightarrow \mu} \left(P_{\nu_{\mu}\rightarrow \nu_{\tau}} {\dfrac{d\Phi}{dE}}^{\!\! \nu_{\mu}} + P_{\nu_{e}\rightarrow \nu_{\tau}} {\dfrac{d\Phi}{dE}}^{\!\! \nu_{e}}\right) + \right. \\
	& \left. \sigma^{cc}_{\bar \nu_{\tau}}(E) Br_{\bar\tau \rightarrow \bar\mu} \left(P_{\bar \nu_{\mu}\rightarrow \bar \nu_{\tau}} {\dfrac{d\Phi}{dE}}^{\!\!\bar \nu_{\mu}} + P_{\bar \nu_{e}\rightarrow \bar \nu_{\tau}} {\dfrac{d\Phi}{dE}}^{\!\!\bar \nu_{e}} \right)\right] ,
\end{aligned}
\end{equation}
where $T$ is the detection time and $Br_{\tau \rightarrow \mu} = Br_{\bar{\tau} \rightarrow \bar{\mu}} \simeq 0.17$ are the branching ratios of the decays $\tau \rightarrow \mu \bar{\nu}_\mu \nu_{\tau}$ and $\bar{\tau} \rightarrow \bar{\mu} \nu_{\mu} \bar{\nu}_{\tau}$. The fluxes of atmospheric neutrinos and antineutrinos were taken from Ref. \cite{Atmnu:1996}. The dependence of ${\mathcal N}_{\mu}$ on the density and composition of the medium is incorporated through the neutrino oscillation probabilities, calculated with the expressions in Sec. \ref{sec:structure_osc} evaluated at a distance $L$. The values of the oscillation parameters are those given in Table \ref{tab:constants} and correspond to the mean of the best-fit values for the allowed ranges at $1\sigma$ of the global analyses implemented by three groups \cite{Capozzi:2020, Esteban:2020cvm, deSalas:2020pgw}.

The charged-current cross sections for $\nu_{\mu}(\bar{\nu}_{\mu})$- nucleon and $\nu_{\tau}(\bar{\nu}_{\tau})$-nucleon scatterings, for neutrino energies in the range 
  $\SI{2}{\giga\electronvolt} < E < \SI{100}{\giga\electronvolt}$, are given approximately by \cite{Formaggio:2013kya}:
\begin{equation}
\begin{aligned}
\sigma^{cc}_{\nu_{\mu}}(E) &\simeq \num{0.75e-38} \left({E}/\unit{\giga\electronvolt}\right) \unit{\square\cm}\,, \\
\sigma^{cc}_{\bar \nu_{\mu}}(E) &\simeq \num{0.35e-38} \left({E}/\unit{\giga\electronvolt}\right) \unit{\square\cm}\,,\\
\sigma^{cc}_{\nu_{\tau}}(E) &\simeq \num{0.13e-38} \left({E}/\unit{\giga\electronvolt}\right) \unit{\square\cm}\,, \\
\sigma^{cc}_{\bar \nu_{\tau}}(E) &\simeq \num{0.05e-38} \left({E}/\unit{\giga\electronvolt}\right) \unit{\square\cm} \,.
\end{aligned} 
\end{equation}  
Non-standard interactions also affect neutrino production and detection through charged-current interactions (CC). In general, the limits for CC-NSI are an order of magnitude more restrictive than for NC-NSI \cite{Davidson2003, Biggio:2009, FMartinez:2011}, and in this work, we disregard their effects on detection.
\begin{figure*}[!ht]
  \centering
  \begin{subfigure}[b]{0.40\textwidth}
    \centering
    \includegraphics[width=\textwidth]{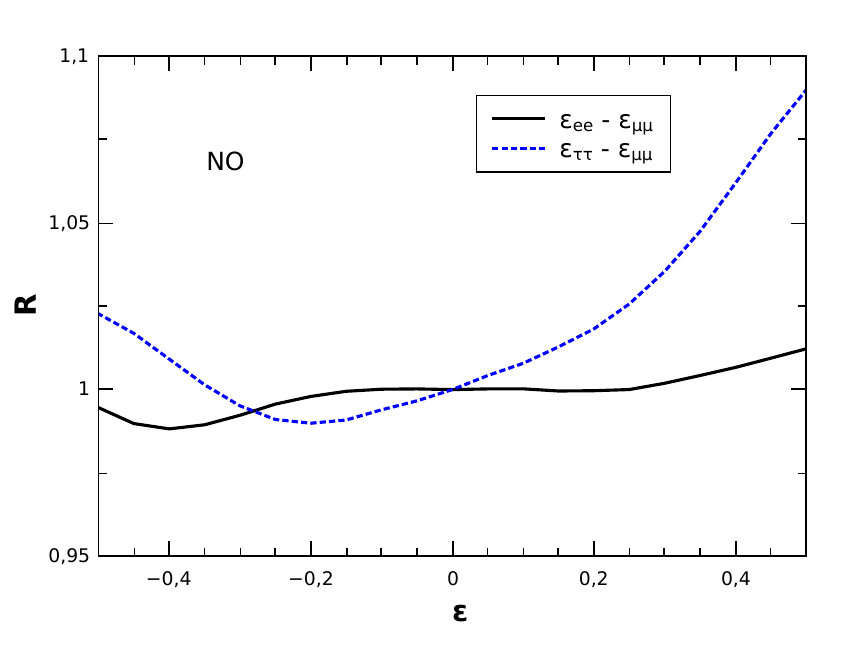}
    \caption{} 
    \label{fig:R_epsNP_diag_NO}
  \end{subfigure}
  \hspace{1cm} 
  \begin{subfigure}[b]{0.40\textwidth}
    \centering
    \includegraphics[width=\textwidth]{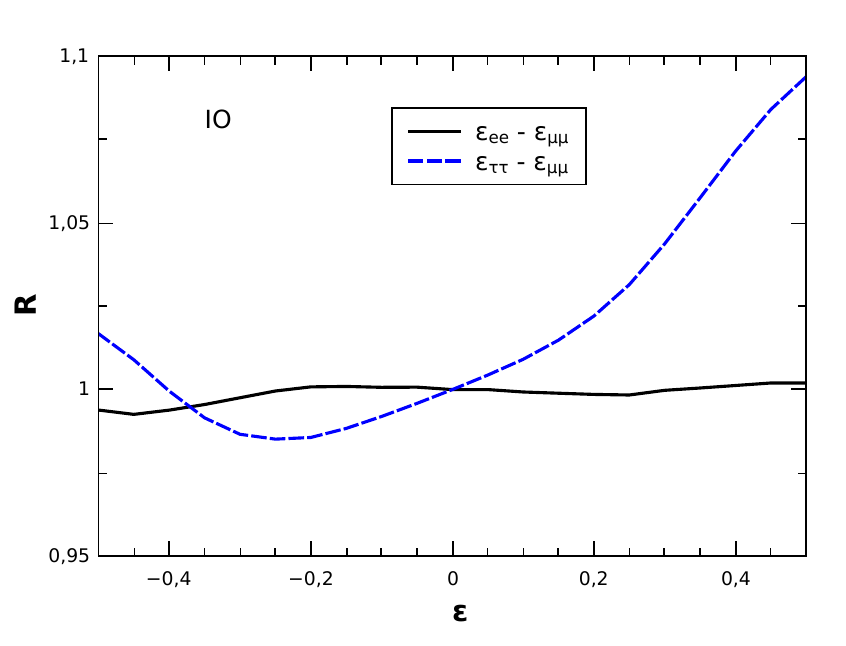}
    \caption{} 
    \label{fig:R_epsNP_diag_IO}
  \end{subfigure}
  \vspace{0.5cm} 
  \begin{subfigure}[b]{0.40\textwidth}
    \centering
    \includegraphics[width=\textwidth]{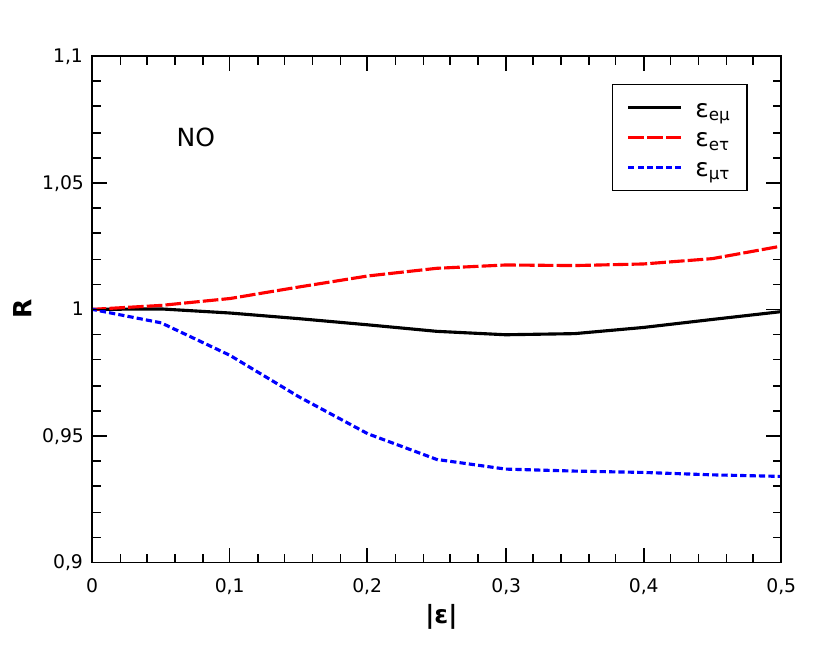}
    \caption{} 
    \label{fig:R_epsNP_nodiag_NO}
  \end{subfigure}
   \hspace{1cm}
  \begin{subfigure}[b]{0.40\textwidth}
    \centering
    \includegraphics[width=\textwidth]{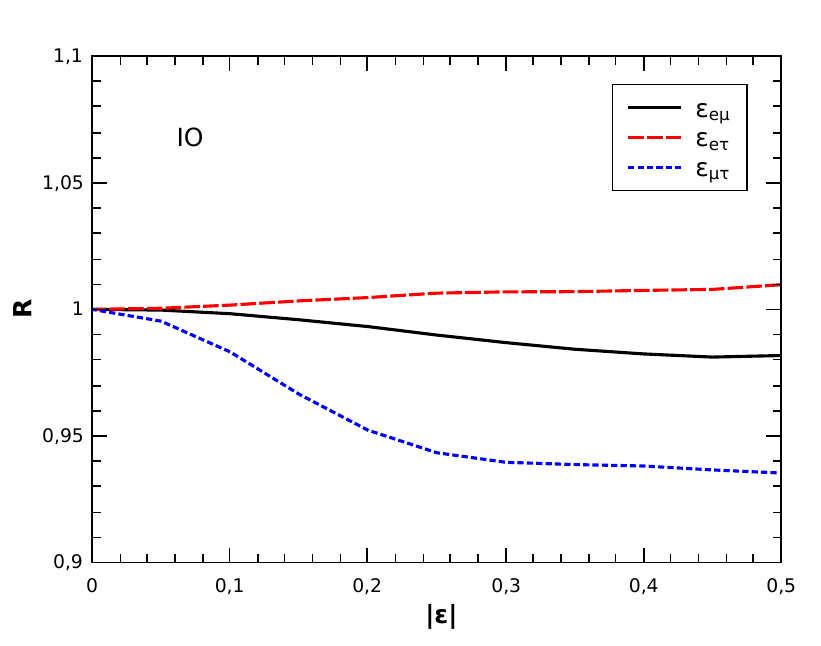}
    \caption{} 
    \label{fig:R_epsNP_nodiag_IO}
  \end{subfigure}
\caption{${R}(\epsilon) = {\mathcal{N}_{\mu}( {\epsilon} )}/{\mathcal{N}_{\mu}( 0 )}$ as a function of a single non-zero entry of the matrix $(\epsilon_{\alpha\beta})$, for the standard Earth and normal and inverted ordering. (a) and (b) Diagonal coefficients $\epsilon_{ee}-\epsilon_{\mu\mu}, \epsilon_{\tau\tau}-\epsilon_{\mu\mu}$. (c) and (d) Modulus of non-diagonal coefficients $\epsilon_{e\mu}, \epsilon_{e\tau}, \epsilon_{\mu\tau}$.}
\label{fig:R_epsNP_nodiag_NO_IO}
\end{figure*}

As shown in Ref. \cite{DOlivo:2020ssf}, the region where $\mathcal{N}_{\mu}$ is most sensitive to changes in the outer core compositions lies in the intervals $\SI{4}{\giga\electronvolt} < E < \SI{10}{\giga\electronvolt}$ and $\ang{10}<\,\eta\,<\ang{60}$ of the neutrinos's energy and nadir angle, respectively. These intervals are contained within the regions for the energies and angles considered in the analysis performed in the following section, which are divided into a grid of $200 \times 200$ square bins. Every pseudo-experiment is made up by tossing in each bin $(i,j)$ of the grid several Poisson-distributed events with the mean value equal to ${\mathcal N}_{\mu}$ as given by Eq.~(\ref{eq:numeroeventos}). Thus, each of the $n_{exp}$ experiments consists of 200 $\times$ 200 numbers corresponding to events, one for each bin. They are called the {\it true events} and assumed to be distributed according to the probability distribution function (pdf) $f^{\,i_{exp}}_t(E,\eta)$, ${i_{exp}} = 1, \ldots, n_{exp}$, given by the normalized histograms constructed using the Monte Carlo simulation.

A realistic distribution of events is implemented by folding the true distribution with the resolution function 
\begin{equation} 
\label{smearing}
\begin{aligned}
\mathcal{S}(E_{\mtt o}, \eta_{\mtt o} \vert E, \eta) = \dfrac{(2\pi)^{-1}}{\Delta \eta \Delta E}\exp\!\left(\!-\dfrac{(\eta - \eta_{\mtt o}\!)^2}{2(\Delta\eta)^2}-\dfrac{(E-E_{\mtt o})^2}{2(\Delta E)^2}\!\right),
\end{aligned}
\end{equation}  
with the detector characterized by the angular and energy resolutions: $\Delta \eta(E) = \alpha_{\eta}/\sqrt{E/\text{GeV}}$ and $\Delta E(E) = \alpha_{E} \, E$, respectively. For the values of the parameters $\alpha_{\eta}$ and $\alpha_{E}$, different situations are possible, as discussed in  Ref. \cite{Rott:2015kwa}. To keep our analysis as simple as possible, we assume a detection efficiency of $100\%$. In terms of the resolution function, the number of {\it observed events} $\mathcal{O}^{\,i_{exp}}_{m,n}$ in the bin $(m,n)$ for the $i_{exp}$-th experiment is given by:
\begin{equation}
\label{observed}
\begin{aligned}
\mathcal{O}^{\,i_{exp}}_{m,n} &= \mathcal{N}_{tot}\sum\limits_{i,j}\;\;\iint\limits_{\text{bin}(m,n)} \! \dd{E}_{\mtt{o}} \dd{\eta}_{\mtt{o}} \; \iint\limits_{\text{bin}(i,j)} \! \dd{E} \, \dd{\eta} \times \\
&\phantom{=} \mathcal{S}(E_{\mtt{o}}, \eta_{\mtt{o}} \vert E, \eta) f^{\,i_{exp}}_t(E, \eta)\,.
\end{aligned}
\end{equation}

\begin{figure*}[!ht]

\vspace{0.5 cm}
  \centering
  \begin{subfigure}[b]{0.38\textwidth}
    \centering
    \includegraphics[width=\textwidth]{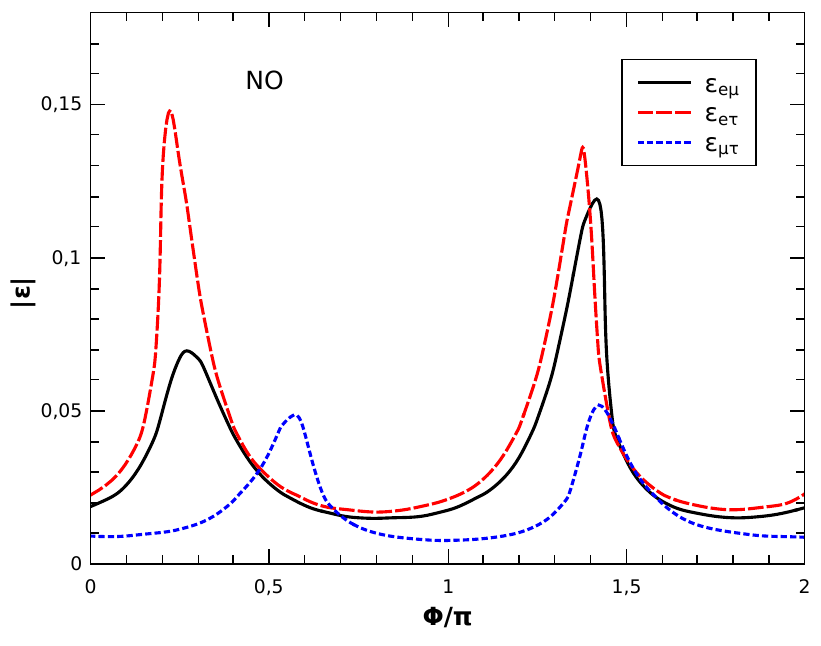}
    \caption{} 
    \label{fig:epsNP_NO}
  \end{subfigure}
  \hspace{1cm} 
  \begin{subfigure}[b]{0.38\textwidth}
    \includegraphics[width=\textwidth]{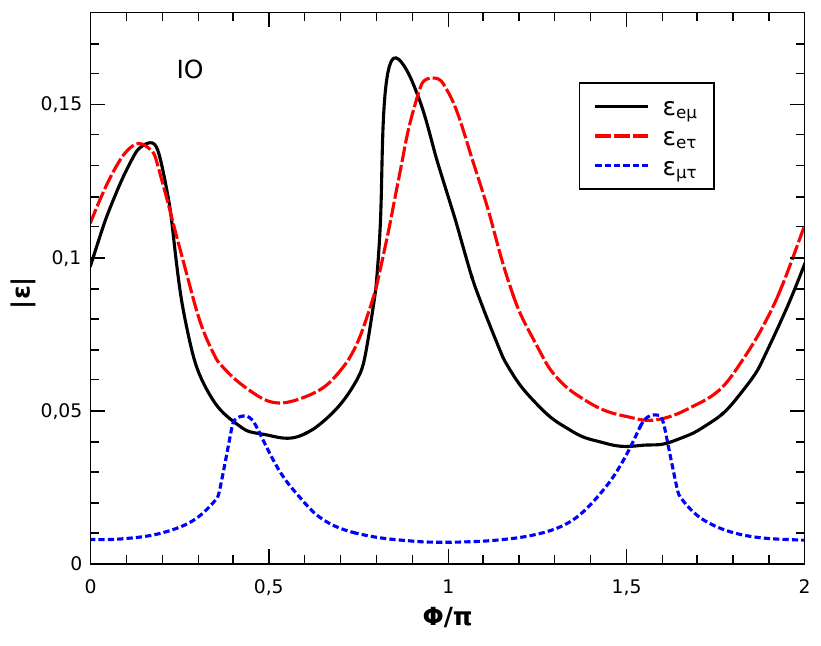}
    \caption{} 
    \label{fig:epsNP_IO}
  \end{subfigure}
  \caption{Bounds on the modulus and phase of the non-diagonal entries of the matrix $(\epsilon_{\alpha\beta})$, for (a) normal ordering and (b) inverted ordering. The allowed values fall in the regions under the curves, where the discrepancy between the events with and without the NSI is less than $1\sigma$.}
  \label{fig:epsNP_NO_IO}
\end{figure*}

As an illustration, in Fig.~\ref{fig:true_obs_NO_IO}, we show the pdf of the observed and true events for the standard Earth defined above and the normal (NO) and inverted ordering (IO) of neutrino masses, respectively, as functions of the energy and nadir angle of atmospheric neutrinos influenced by their SM weak interactions with matter,  as shown in \cite{DOlivoSaez:2022vdl}. The figures were made using the same (large) number of bins for both kinds of events, but these numbers generally differ. In what follows, to test how well the hypothesis of different Earth compositions agrees with the standard Earth, we will distribute the observed events into five angular bins and nine energy bins. 

From Eq. \eqref{observed}, for each bin $(m,n)$, with  $m = 1, \ldots, 9$ and $n = 1, \ldots, 5$, we determine the number of events $\mathcal{O}^{\hspace{0.01 cm}{0}\,i_{exp}}_{\alpha}$ for the standard Earth and SM interactions, as well as for $ \mathcal{O}^{{\hspace{0.02 cm}i_{exp}}}_{\alpha}\big((Z/A)_{oc}, \epsilon\big)$ 
for an alternative Earth, with a different composition $(Z/A)_{oc}$ of the outer core, and NSI coefficients, generically denoted by $\epsilon$. Here, $\alpha = 1, \ldots , 45$ label the two-dimensional bins ($m,n$). Thus, for each of these bins, we have a sample of the observed events and determine the corresponding mean values
\begin{equation}
\label{eq:barOalpha}
\mathcal{\bar{O}}^{\mtt{0}}_{\alpha} = \dfrac{1}{n_{exp}}\sum_{i_{exp}}\mathcal{O}^{\hspace{0.01 cm}{\mtt{0}}\,i_{exp}}_{\alpha}, \hspace{1 cm}
\mathcal{\bar{O}}_{\alpha} = \frac{1}{n_{exp}}\sum_{i_{exp}}\mathcal{O}^{{\hspace{0.02 cm}i_{exp}}}_{\alpha}.
\end{equation}
For Poisson distributed events, in terms of the likelihood function $L$, we construct the negative log-likelihood ratio function as
\begin{equation}
\label{eq:chi2lambda}
\begin{aligned}
\chi^2_{\lambda} &= - 2\ln\left[\frac{L\left(\mathcal{\bar{O}}^{\mtt{0}}_{\alpha};\mathcal{\bar{O}}_{\alpha}\right)}{L\left(\mathcal{\bar{O}}^{\mtt 0}_{\alpha},\mathcal{\bar{O}}^{\mtt{0}}_{\alpha}\right)}\right]\\
&= 2\sum_{\alpha}\!\left[\mathcal{\bar O}_{\alpha}-\mathcal{\bar{O}}^{\mtt{0}}_{\alpha} + \mathcal{\bar O}^{\mtt{0}}_{\alpha} \ln\left(\frac{\mathcal{\bar{O}}^{\mtt{0}}_{\alpha}}{\mathcal{\bar{O}}_{\alpha}}\right)\!\right].
\end{aligned}
\end{equation}
For simplicity, when writing Eqs. \eqref{eq:barOalpha} and \eqref{eq:chi2lambda}, we omitted the dependencies of $\mathcal{O}^{{\hspace{0.02 cm}i_{exp}}}_{\alpha}$ and $\mathcal{\bar O}_{\alpha}$ on the quantities $(Z/A)_{oc}$ and $\epsilon$. 
According to Wilks' theorem \cite{wilks1938}, the $\chi^2_{\lambda}$ distribution can be approximated by the $\chi^2$ distribution and, from it, the goodness of the fit can be established. The statistical significance of the $\chi^{2}$ test is given, as usual, by the $\mathbf{p}$-value: 
\begin{equation}
\mathbf{p} = \int_{\chi^2}^\infty \! f_{\chi}(w,n_{_{\!\rm dof}}) \dd{w}\,,
\end{equation}
where $n_{_{\!\rm dof}}$ is the number of degrees of freedom and $f_{\chi}(w,n_{_{\!\rm dof}})$ is the chi-square distribution. Here, $n_{_{\!\rm dof}} = 9\times5 -2 = 43$. 

From the above expressions, we can examine the levels of discrepancy between the standard Earth and different hypotheses about the composition of the outer core. This is addressed in the next section, paying attention to how it might be affected by new physical effects in the neutrino sector

\begin{figure*}[!ht]
  \centering
  \begin{subfigure}[b]{0.40\textwidth}
    \centering
    \includegraphics[width=\textwidth]{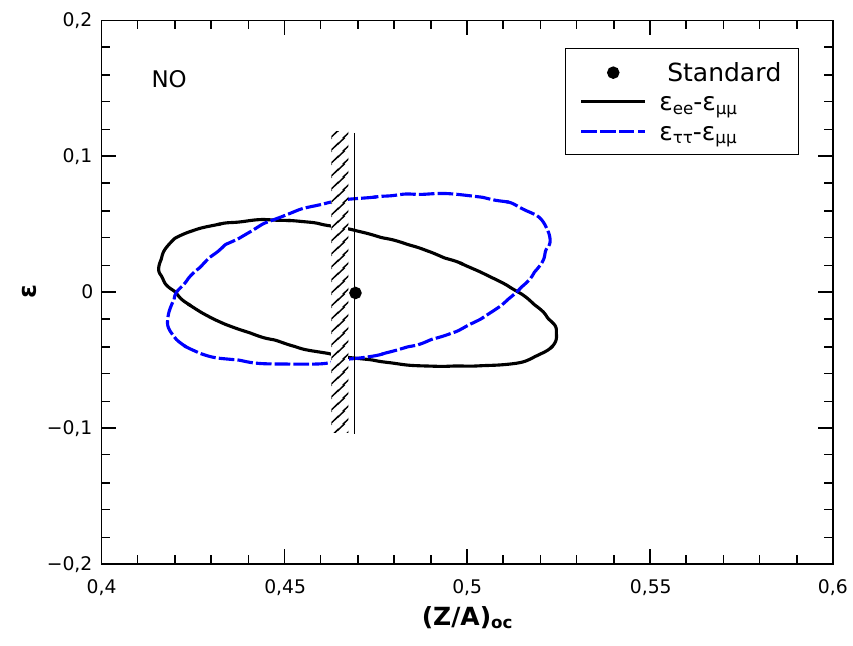}
    \caption{Diagonal coefficients, normal ordering.}
    \label{fig:epsNP_diag_c1_NO}
  \end{subfigure}
  \hspace{1cm} 
  \begin{subfigure}[b]{0.40\textwidth}
    \centering
    \includegraphics[width=\textwidth]{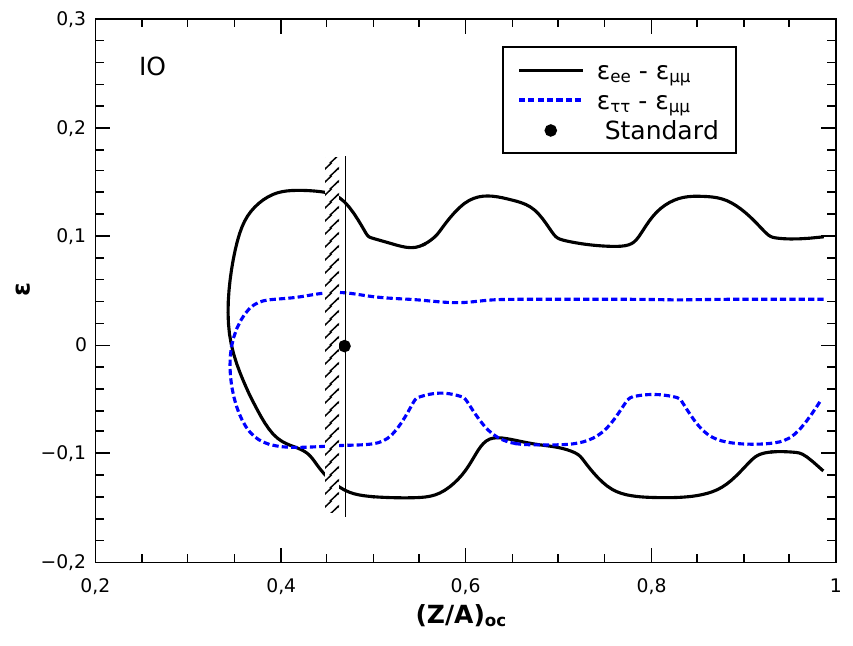}
    \caption{Diagonal coefficients, inverted ordering.}
    \label{fig:epsNP_diag_c1_IO}
  \end{subfigure}
  \vspace{0.5cm} 
  \begin{subfigure}[b]{0.40\textwidth}
    \centering
    \includegraphics[width=\textwidth]{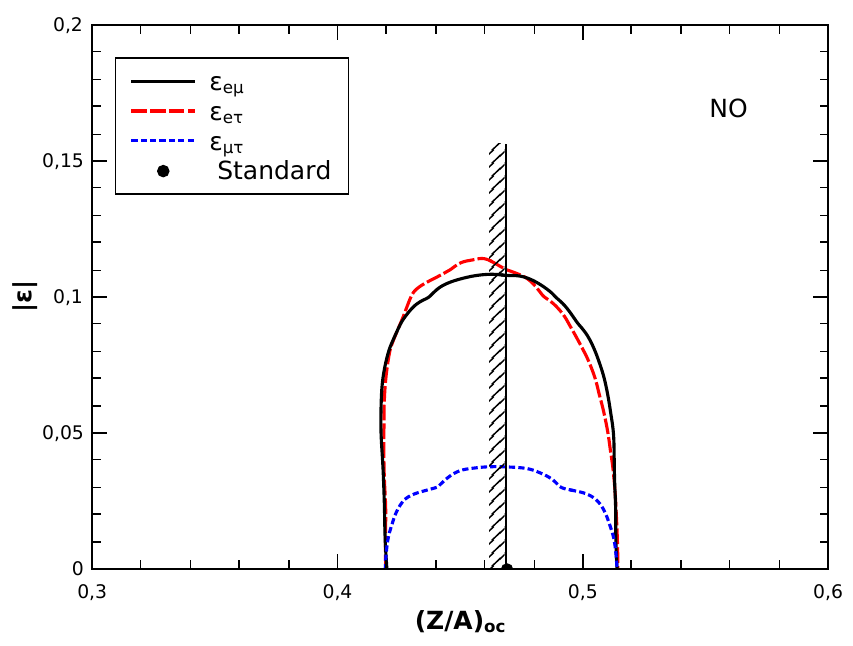}
    \caption{Non-diagonal coefficients, normal ordering.}
    \label{fig:epsNP_nodiag_c1_NO}
  \end{subfigure}
  \hspace{1cm}  
  \begin{subfigure}[b]{0.40\textwidth}
    \centering
    \includegraphics[width=\textwidth]{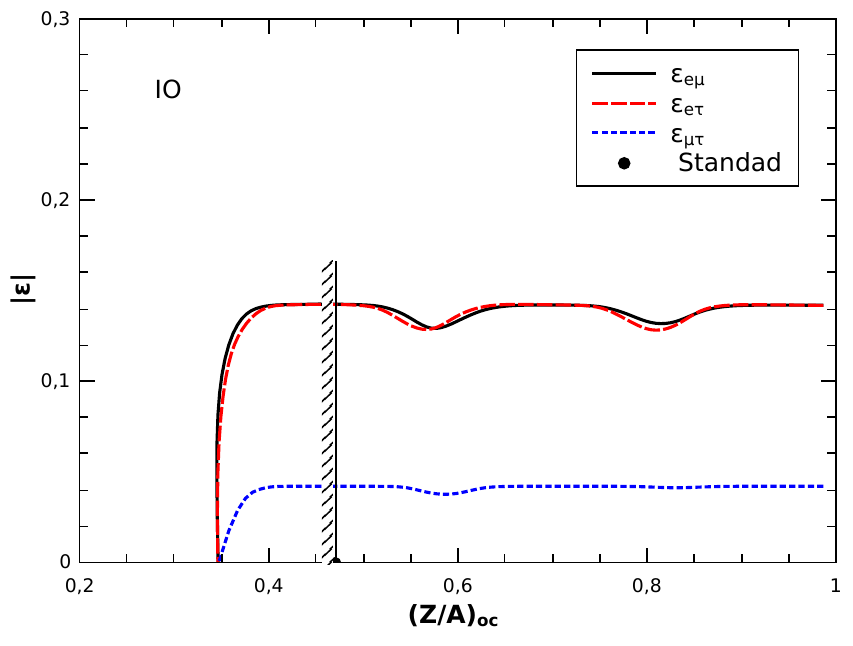}
    \caption{Non-diagonal coefficients, inverted ordering.}
    \label{fig:epsNP_nodiag_c1_IO}
  \end{subfigure}
  \caption{Entries of the matrix $(\epsilon_{\alpha\beta})$ as a function of the outer core composition. Entries are considered nonzero, one at a time. (a) and (b) Diagonal coefficients $\epsilon_{ee} - \epsilon_{\mu\mu}$, $\epsilon_{\tau\tau} - \epsilon_{\mu\mu}$. (c) and (d) Non-diagonal coefficients $\epsilon_{e\mu}$, $\epsilon_{e\tau}$, $\epsilon_{\mu\tau}$, with the respective phases marginalized. In the regions enclosed by the contours, the difference between events with non-standard $(Z/A)_{oc}$ and/or NSI and those for the standard Earth without NSI (the black dot in the figures) is less than a significance level of $1\sigma$, for normal and inverted mass ordering. The regions to the right of the vertical dashed line correspond to the standard composition plus the addition of hydrogen in the outer core.}
  \label{fig:compo_eps}
\end{figure*}


\section{Tomography of the Earth's core in the presence of NC-NSI}
\label{sec:results}

\begin{figure*}[!ht]
  \centering
  \begin{subfigure}[b]{0.40\textwidth}
    \centering
    \includegraphics[width=\textwidth]{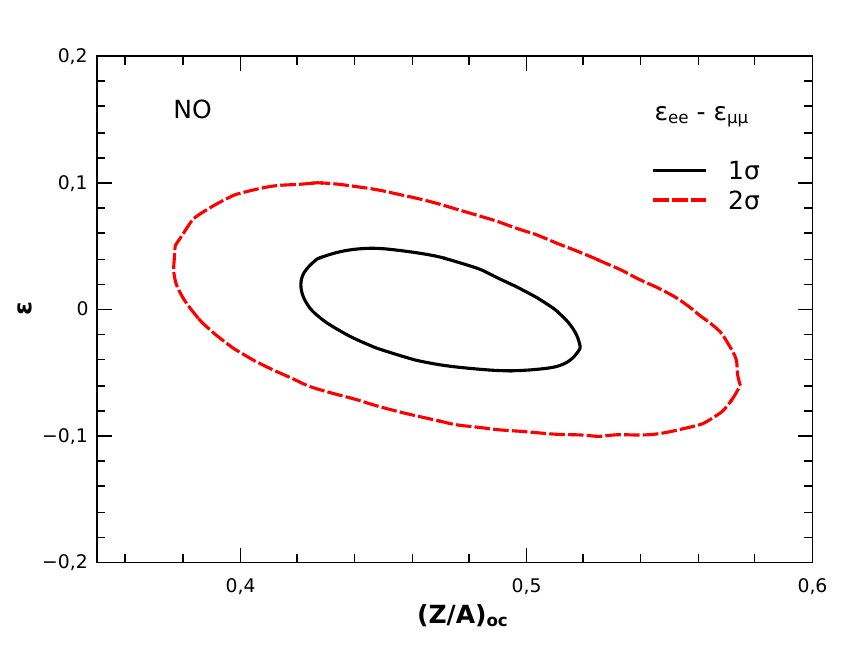}
    \caption{Diagonal coefficient $\epsilon_{ee} - \epsilon_{\mu\mu}$, normal ordering.}
    \label{fig:eps_ee_NO}
  \end{subfigure}
  \hspace{1cm} 
  \begin{subfigure}[b]{0.40\textwidth}
    \centering
    \includegraphics[width=\textwidth]{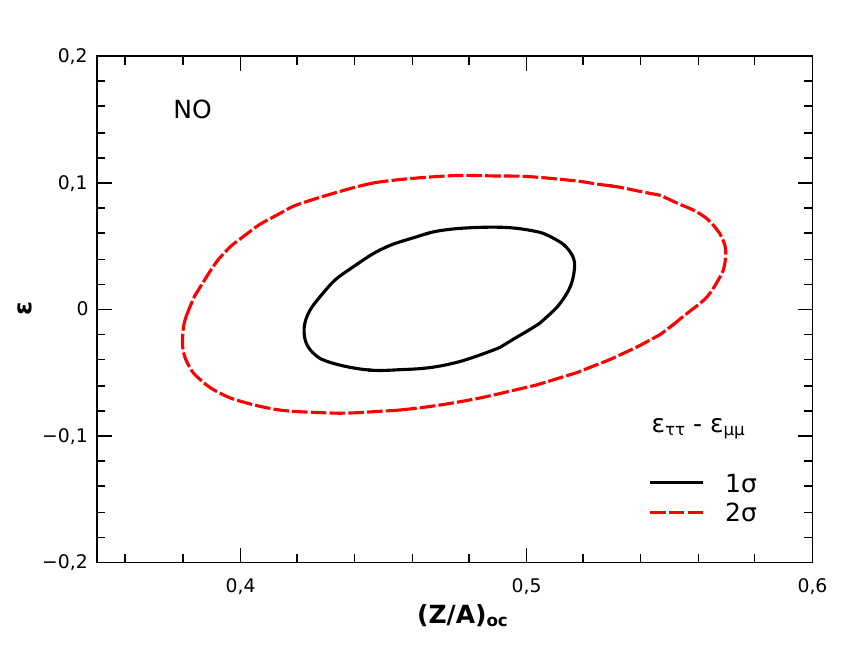}
    \caption{Diagonal coefficient $\epsilon_{\tau\tau} - \epsilon_{\mu\mu}$, normal ordering.}
    \label{fig:eps_tt_NO}
  \end{subfigure}
  \caption{Diagonal coefficients, (a) $\epsilon_{ee} - \epsilon_{\mu\mu}$ and (b) $\epsilon_{\tau\tau} - \epsilon_{\mu\mu}$, as functions of the outer core composition, for normal ordering. The regions enclosed by the contours correspond to the $1\sigma$ (continuous line) and $2\sigma$ (dashed line) statistical significance levels of the difference between events with non-standard $(Z/A)_{oc}$ and/or NSI and those for the standard Earth without NSI.}
  \label{fig:2-3sigmacorrelations}
\end{figure*}

\begin{figure*}[!ht]
\centering
\begin{subfigure}[b]{0.40\textwidth}
\includegraphics[width=\textwidth]{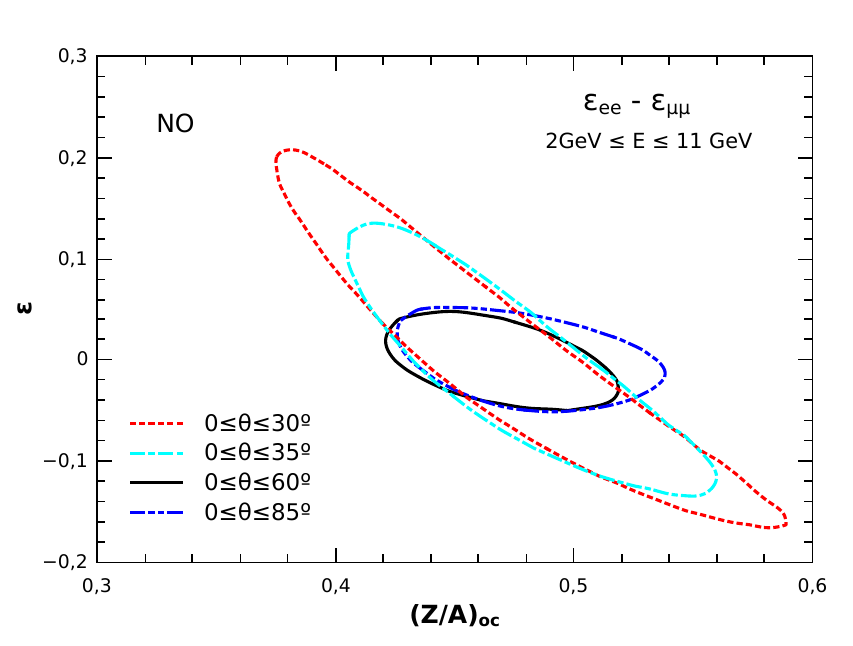}
\caption{}	
\label{fig:ee_ang}	
\end{subfigure}
\hspace{1cm} 
\begin{subfigure}[b]{0.40\textwidth}
 \centering
\includegraphics[width=\textwidth]{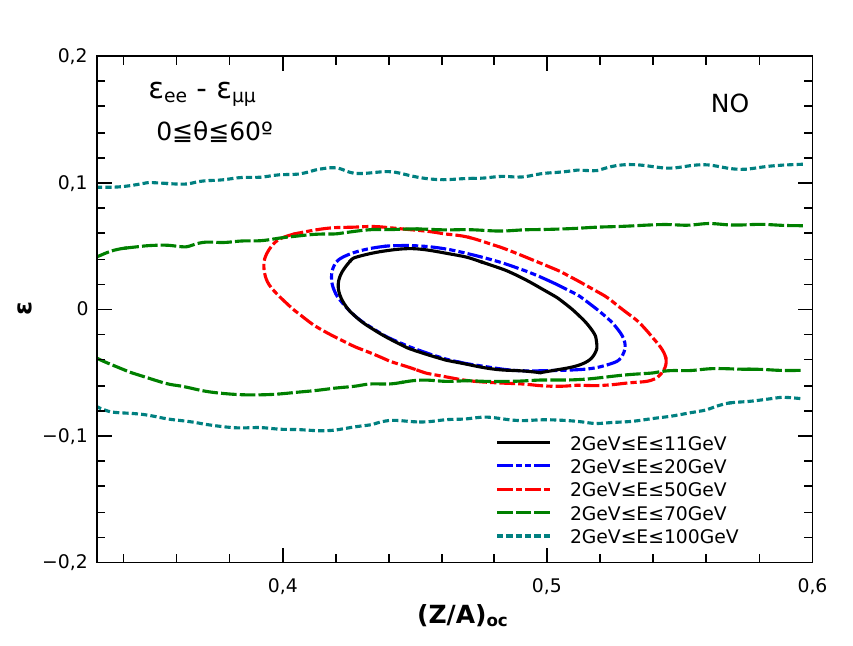}
\caption{}	
\label{fig:ee_ene}	
\end{subfigure}
\vspace{0.5cm} 
\begin{subfigure}[b]{0.40\textwidth}
\includegraphics[width=\textwidth]{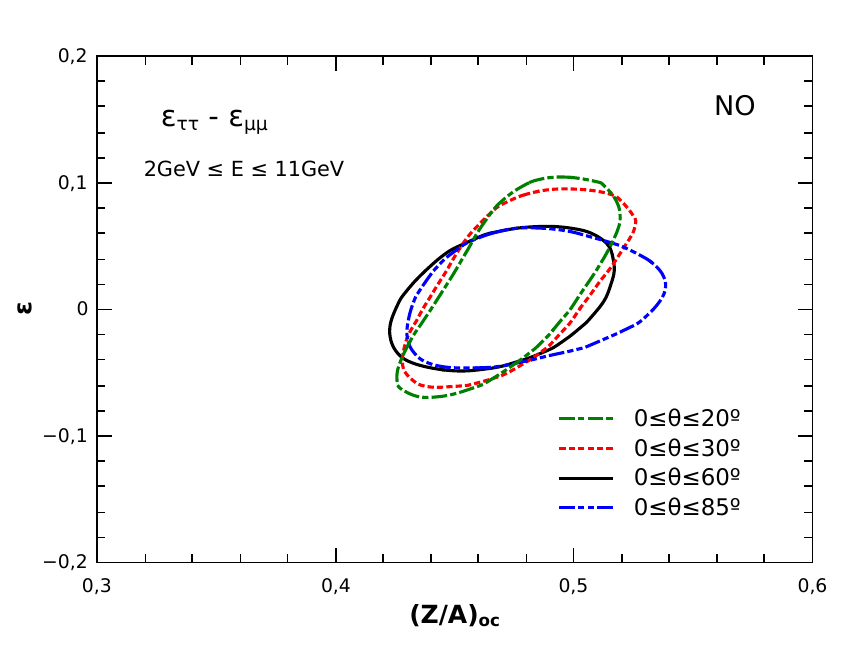}	
\caption{}
\label{fig:tt_ang}
\end{subfigure}
  \hspace{1cm} 
\begin{subfigure}[b]{0.40\textwidth}
\includegraphics[width=\textwidth]{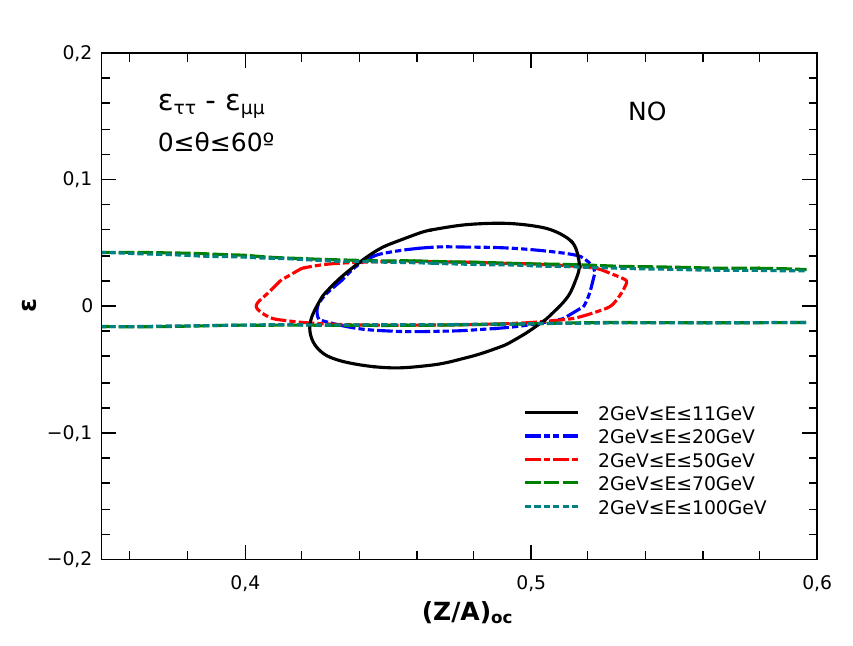}	
\caption{}
\label{fig:tt_ene}
\end{subfigure}
\caption{1$\sigma$ regions for the diagonal NSI coefficients, for different angular intervals and different energy intervals\\
(a) and (b) $\epsilon_{ee}-\epsilon_{\mu\mu}$\,. (c) and (d) $\epsilon_{\tau\tau}-\epsilon_{\mu\mu}$\,.}
\label{fig:1sigregion_ang&energy}
\end{figure*}

To visualize the behavior of the events as a function of the coefficients $\epsilon_{\alpha \beta}$, we introduce the quantity
\begin{equation}
\label{eq:eRe}
{R}(\epsilon) \equiv \dfrac{\mathcal{N}_{\mu}( {\epsilon} )}{\mathcal{N}_{\mu}( 0 )}\,,
\end{equation}
which gives the relative size between the number ${\mathcal{N}_{\mu}( {\epsilon})}$ of $\mu$ events calculated including the effect of the different 
NC-NSI entries of the potential energy matrix, and the number $\mathcal{N}_{\mu}(0)$ calculated for the standard interactions only.   
In Fig.~\ref{fig:R_epsNP_nodiag_NO_IO}, we show ${R}(\epsilon)$ as a function of a single non-zero entry of $(\epsilon_{\alpha \beta})$. The events 
are calculated in the intervals $\SI{2}{\giga\electronvolt} < E < \SI{11}{\giga\electronvolt}$ and $\ang{0}<\,\eta\,<\ang{60}$, for NO and IO of neutrino masses. 
This is done for the standard Earth and the oscillation parameters in Table \ref{tab:constants}, assuming a 10-year operation of an 8 Mton detector like ORCA, 
with resolution parameters $\alpha_{\eta} = 0.25$ and $\alpha_{E} = 0.20$. 

\begin{figure*}[!htbp]
  \centering
  \begin{subfigure}[b]{0.40\textwidth}
    \centering
    \includegraphics[width=\textwidth]{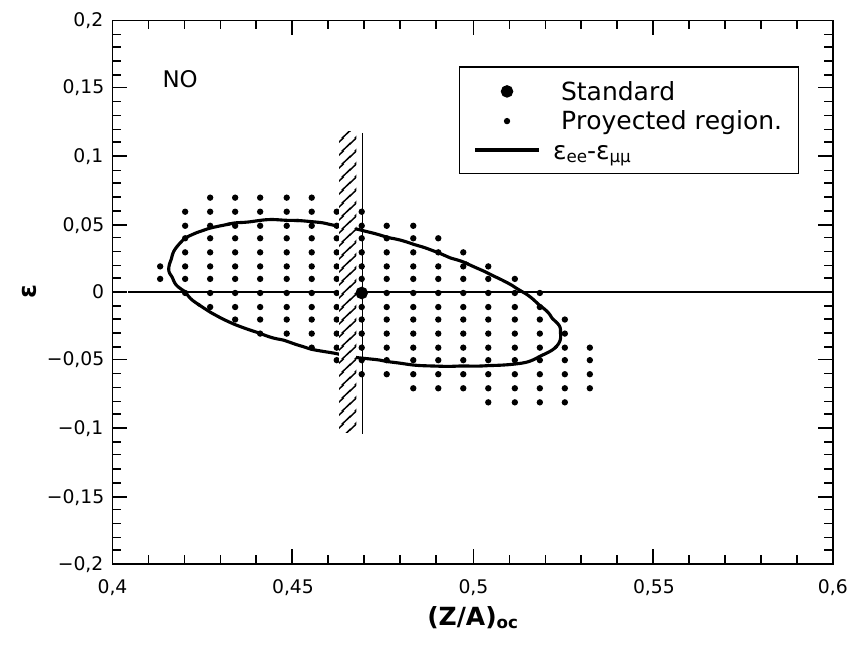}
    \caption{Diagonal coefficient $\epsilon_{ee} - \epsilon_{\mu\mu}$, normal ordering.}
    \label{fig:eps_ee_NO_density}
  \end{subfigure}
  \hspace{1cm}  
  \begin{subfigure}[b]{0.40\textwidth}
    \centering
    \includegraphics[width=\textwidth]{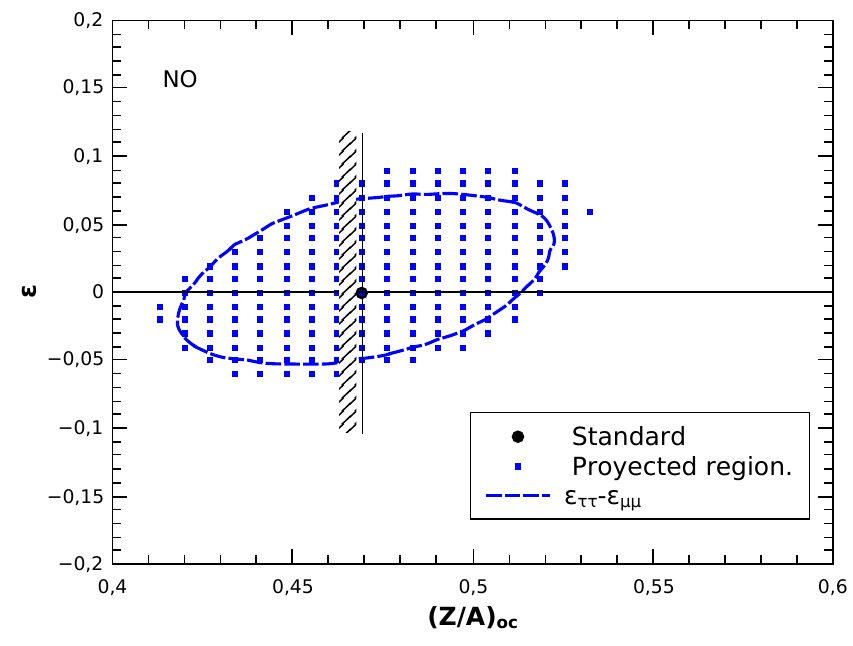}
    \caption{Diagonal coefficient $\epsilon_{\tau\tau} - \epsilon_{\mu\mu}$, normal ordering.}
    \label{fig:eps_tt_NO_density}
  \end{subfigure}
  \caption{The same regions as Figs.~\ref{fig:epsNP_diag_c1_NO} and \ref{fig:epsNP_diag_c1_IO}, but including uncertainties in the outer core density, for normal ordering. The shaded areas correspond to the projections onto the plane $\big((Z/A)_{oc}, \epsilon\big)$ of the $1\sigma$ confidence volume obtained by fitting these quantities together with the relative change of the outer core density $\delta_{oc}$.}
  \label{fig:densityproyections}
\end{figure*}
The variables to be adjusted are the composition of the outer core and the magnitudes of the NSI coefficients. To simplify the analysis as much as possible, the coefficients are taken to be nonzero, one at a time. As a first look at their possible allowed values, using the same procedure described in the previous paragraph, for the non-diagonal entries we construct regions in the plane \,($|\epsilon_{\alpha \beta}|, \phi_{\alpha \beta}/\pi$), $\al \neq \beta$, where $\phi_{\alpha \beta}\!=\hsn$ arg\,$\epsilon_{\alpha \beta}$. In Fig.~\ref{fig:epsNP_NO_IO}, we show the limits for the modulus and phase of the complex quantities $\epsilon_{e\tau}, \epsilon_{e\mu}$, and\,$\epsilon_{\mu\tau}$, for NO and IO. The allowed values fall in the regions under the curves, where the statistical significance of the discrepancy between the events with and without the NSI is less than a significance level of $1\sigma$. 

In the following, to typify how the presence of NSI can affect tomographic studies of the outer core composition $(Z/A)_{oc}$, we construct regions in the plane $\left((Z/A)_{oc}, \epsilon_{\alpha \beta}\right)$. Figures \ref{fig:epsNP_diag_c1_NO} and \ref{fig:epsNP_diag_c1_IO}, show graphs of the real diagonal entries $\epsilon_{ee} - \epsilon_{\mu\mu}$ and $\epsilon_{\tau\tau} - \epsilon_{\mu\mu}$ as functions of the outer core composition, while Figures \ref{fig:epsNP_nodiag_c1_NO} and \ref{fig:epsNP_nodiag_c1_IO} show the corresponding graphs for the modules of the complex off-diagonal entries 
$\epsilon_{e\mu},\epsilon_{e\tau},$ and $\epsilon_{\mu\tau}$, with the respective phases marginalized. The allowed values lie in the regions inside the curves, where again, the statistical significance level of the difference between events with non-standard $(Z/A)_{oc}$ and/or NSI and those for the standard Earth without NSI (the black dot in the figures) is less than $1\sigma$.  As can be seen, the regions for NO are considerably more restrictive than those for IO. In the case of the diagonal coefficients, there appears to be a sort of negative (positive) correlation for $\epsilon_{ee} - \epsilon_{\mu\mu}$ ($\epsilon_{\tau\tau} - \epsilon_{\mu\mu}$), which would indicate that an increase (decrease) in composition is offset by a decrease (increase) in the NSI coefficient, and vice versa. This effect is similar to the compensation found for changes in the composition and density of the outer core \cite{DOlivoSaez:2022vdl}, and may be associated with the fact that the potential energy in the Hamiltonian of Eq. \eqref{hamiltonian} depends on the product of the composition and the NSI coefficient.  Such compensation is partial, since changes in matter effects associated with variations in the NSI coefficients affect neutrino oscillations in the other regions of the Earth as well, not just in the outer core. Consequently, the allowed regions in the plane $\big((Z/A)_{oc}, \epsilon\big)$ become closed. We verified that such correlations persist at $2 \sigma$, as shown in Fig. \ref{fig:2-3sigmacorrelations}.  No correlations are observed for non-diagonal entries.

While the work reported here does not focus on quantifying the knowledge of the size and flavor structure of the neutral current non-standard neutrino interactions, it is worth noting that the resulting constraints for the NSI coefficients are compatible with those bounds reported in the literature, derived from global analysis of $3\nu$ oscillation data \cite{Esteban:2019, Coloma:2023ixt} and those reported by experiments (See \cite{KM3NeT:2024pte} and references therein). In the global analyses, all parameters (oscillation and non-standard) are varied simultaneously, and the sensitivity to each of them is obtained by marginalizing the remaining that are not shown in the corresponding graphs. Since energies higher than those in our analysis are accessible to current and upcoming experiments, such as IC-DeepCore, Super-Kamiokande, and KM3NeT/ORCA, it is valid to ask if modifying energy and nadir angle ranges might enhance sensitivity to NSIs and help resolve degeneracies related to core composition. With this in mind, we consider higher energies and different nadir angle intervals. The results for the diagonal coefficients and NO are presented in Fig. \ref{fig:1sigregion_ang&energy}.  As the nadir angle increases, the relative contribution of the core decreases, the lengths of the paths through it decrease, and for angles greater than $30^{\circ}$, some paths don't even cross it. This results in a reduced ability to determine the composition of the outer core, as best observed for a vanishing NSI coefficient in Figs. \ref{fig:ee_ang} and \ref{fig:tt_ang}, for $\SI{2}{\giga\electronvolt} < E < \SI{11}{\giga\electronvolt}$. Conversely, as the angle decreases, the contribution of the core increases, and with it, the compensation effect mentioned above. Accordingly,  the size of the $1 \sigma$ region grows, which is more evident in the case of $\epsilon_{ee} - \epsilon_{\mu\mu}$. Regarding the neutrino energy, as its range becomes larger, the determination of the composition of the outer core worsens, but in this case, the allowed interval increases for $\epsilon_{ee} - \epsilon_{\mu\mu}$ increases but theone for $\epsilon_{\tau\tau} - \epsilon_{\mu\mu}$ decreases. 
From the inspection of the different graphs in Fig. \ref{fig:1sigregion_ang&energy}, it is evident that the region delimited by the black line, corresponding to the intervals 
$\SI{2}{\giga\electronvolt} < E < \SI{11}{\giga\electronvolt}$ and $\ang{0}<\,\eta\,<\ang{60}$, is the one in which the best simultaneous fit of the outer core composition and the magnitude of the diagonal NSI coefficients is obtained.
\begin{figure*}[!htbp]
  \centering
  \begin{subfigure}[b]{0.40\textwidth}
    \centering
    \includegraphics[width=\textwidth]{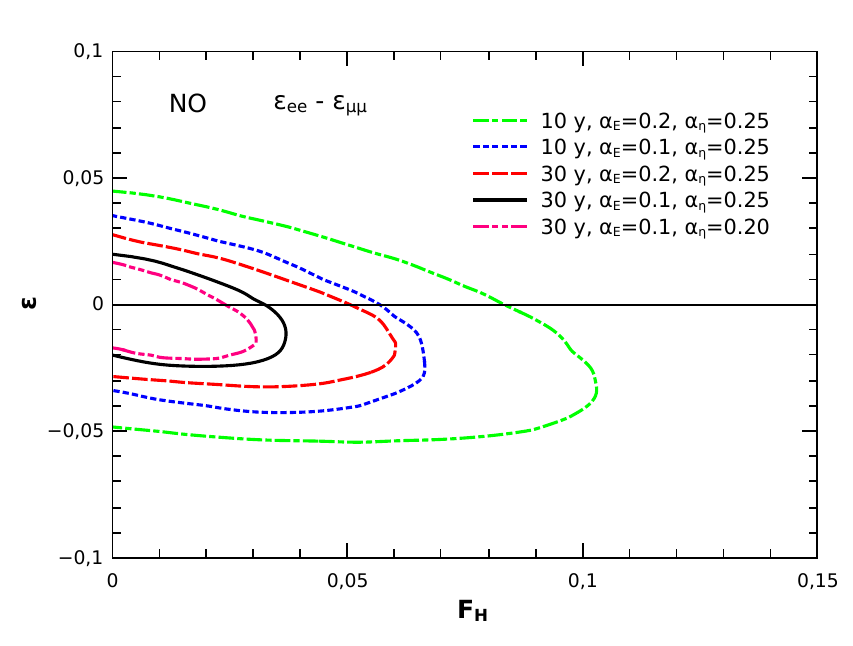}
    \caption{}
    \label{fig:thirtyyearsexpositions_a}
  \end{subfigure}
  \hspace{1cm}  
  \begin{subfigure}[b]{0.40\textwidth}
    \centering
    \includegraphics[width=\textwidth]{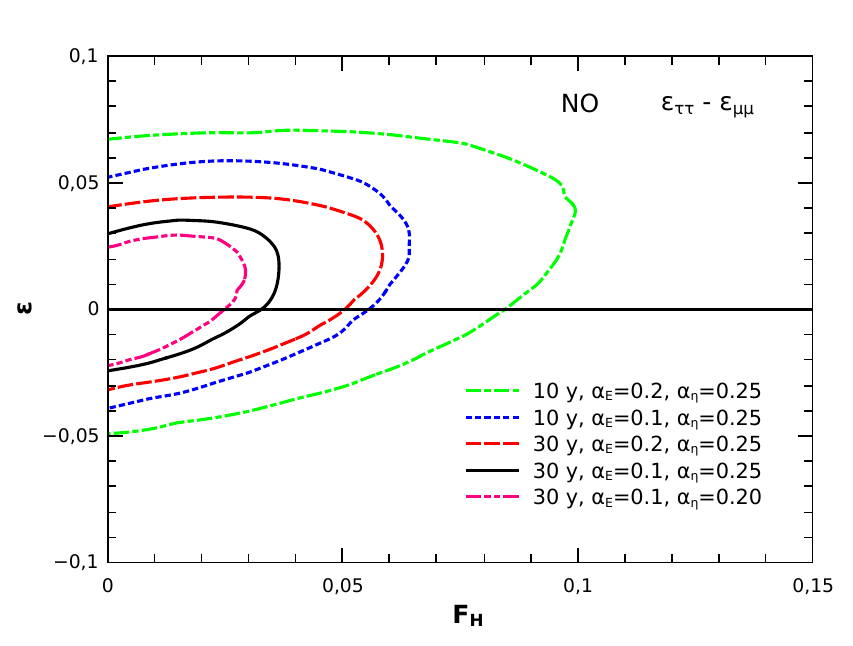}
  \caption{}
    \label{fig:thirtyyearsexpositions_b}
  \end{subfigure}
  
  \vspace{0.5cm} 
  \begin{subfigure}[b]{0.40\textwidth}
    \centering
    \includegraphics[width=\textwidth]{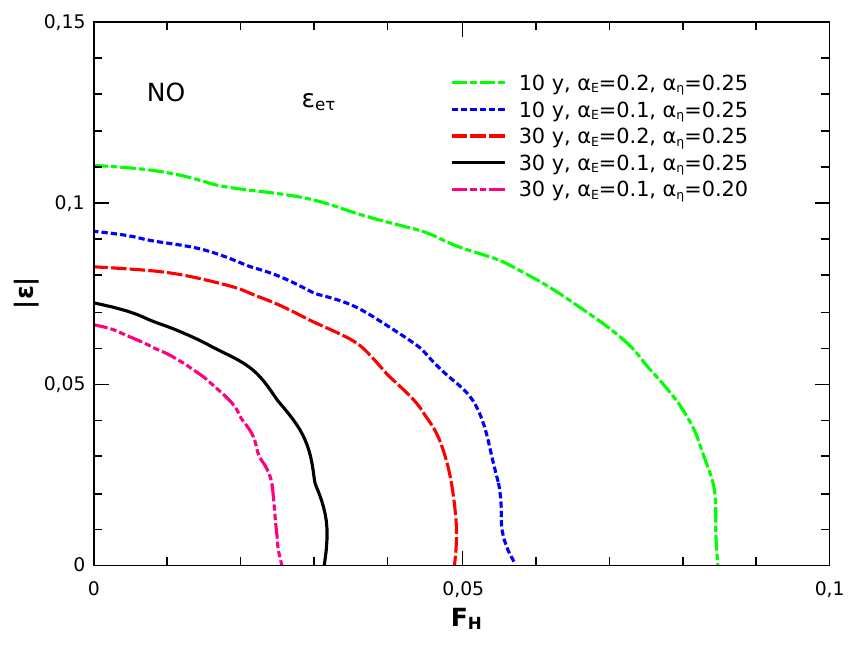}
    \caption{}
    \label{fig:thirtyyearsexpositions_c}
  \end{subfigure}
  \caption{Hydrogen fraction in the outer core for different exposures and detector resolutions, for normal ordering of neutrino masses.
(a) $\epsilon_{ee}-\epsilon_{\mu\mu}$, (b) $\epsilon_{\tau\tau}-\epsilon_{\mu\mu}$, and (c) $\epsilon_{e\tau}$.}
  \label{fig:thirtyyearsexpositions}
\end{figure*}

For terrestrial parameters, we retain the unadjusted PREM values, except when, to account for the effects that uncertainties in matter densities can introduce into compositional fits, we add the relative change in outer core density $\delta_{oc}$. In doing so, it must be noted that the densities of the lower and upper mantle have to change proportionally to $\delta_{oc}$ to satisfy the constraints on the sum of the masses and the sum of the moments of inertia of the Earth's main layers set out in Section 2 \cite {DOlivoSaez:2022vdl}. The density changes translate into modifications of the electron number density along the neutrino path, which, in turn, affect the probability of flavor transitions. Consequently, fitting  $(Z/A)_{oc}$ and the NSI coefficient together with $\delta_{oc}$ gives a 1$\sigma$ volume 
which, when projected onto the $\big((Z/A)_{oc}, \epsilon\big)$, results in the shaded areas in Fig.~\ref{fig:densityproyections}.

\section{Hydrogen content and final comments}

As we have pointed out, our goal is to understand to what extent the presence of hydrogen in the outer core of our planet can be detected by the modifications it introduces into the flavor oscillations of atmospheric neutrinos, when possible NC-NSI effects are not ignored. In the graphs of Fig. \ref{fig:compo_eps}, the regions to the right of the vertical dashed line correspond to the standard composition plus the addition of hydrogen in the outer core.  Assuming that the variation in the composition of the outer core is associated exclusively with the abundance of hydrogen, we can write
\be
(Z/A)_{oc}=(1-\digamma_{\!\!\chic H})(Z/A)_{oc}^{\chic 0} + \digamma_{\!\!\chic H}(Z/A)^{\chic H}_{oc}\,,
\ee
where $(Z/A)^{\chic 0}_{oc} = 0.4691$ is the value with no hydrogen and $\digamma_{\!\!\chic H} = M_{\chic H}/M_{oc}$ is the fractional contribution that hydrogen makes to the total mass of the layer. Without NSI, for NO, the 1$\sigma$ intervals are compatible with $\sim 8$ wt\% of hydrogen in the outer core, a value too high in light of geophysical estimates. Therefore, an experiment like the one we consider here clearly has a rather limited potential to reveal the presence of lighter elements in the Earth's core simultaneously with NC-NSI interactions. This worsens for the inverse mass ordering. 

Then, assuming that neutrino oscillations were affected by non-standard interactions, it is pertinent to ask what exposure and/or detector resolutions would be required to significantly constrain the presence of hydrogen using an oscillation tomography study of the deep regions of our planet.  As a partial answer, in Fig.~\ref{fig:thirtyyearsexpositions}, we show the 1$\sigma$ regions for two different exposures and two energy resolutions. It can be seen that with a 30-year exposure, it is possible to constrain the hydrogen content to 5 $\%$ and 3 $\%$ with energy resolutions of 0.20 and 0.10, respectively. From the graphs of the diagonal coefficients, it is also clear that the extreme values for the hydrogen fraction are only feasible for large values ($|\epsilon| \simeq 0.05$) of the coefficients. To exemplify the impact of an improvement in angular resolution, we also include a graph of 30 years of data collection, with $\alpha_{E} = 0.10$ and $\alpha_{\eta} = 0.20$. On the other hand, assuming the value $(Z/A)^{\chic 0}_{oc}$ (i.e., $\digamma_{\!\!\chic H}=0$) for the outer core composition, we observe that, for ten years of operation with a sensitivity of 0.20, the NSI coefficients are restricted approximately as follows: $\epsilon_{ee} - \epsilon_{\mu\mu}\in[-0.047, 0.045]$, $\epsilon_{ee} - \epsilon_{\tau \tau}\in[-0.049, 0.068]$, and $\vert\epsilon_{e\tau}\vert \leq 0.11$. The first of these restrictions coincides with the one obtained in Ref. \cite{Denton:2021rgt}, when considering the sensitivity of the incoming experiment DUNE to the matter effect in the Earth rescaled by $1+\epsilon_{ee}$.

In summary, the next generation of neutrino detectors could detect significant quantities of hydrogen in the Earth's deep internal regions through the neutrino oscillation tomography of the planet. They will also be able to investigate physics beyond the SM, usually described in a model-independent manner by NSI. In this work, we consider the combined effects of uncertainties in the composition of the outer core and NC-NSI on the propagation of atmospheric neutrinos within terrestrial matter. Our goal was to examine the interplay between these two effects: to what extent NSI can hinder the acquisition of compositional information and, vice versa, how compositional uncertainties could mask new physics effects. Using the $\mu$-type events produced in a generic large Cherenkov detector as physical observables and performing a Monte Carlo simulation of the energy and azimuthal angle distribution of these events, we test possible variations in the composition of the outer core together with the effects of new interactions in the neutrino sector.  Considering neutrino oscillations in a standard Earth, without H in the core, we constrain the possible values of the dimensionless coefficients $\epsilon_{\al \beta}$ that parametrize the NSI intensities relative to the SM interactions. Then, for the normal-ordering and inverted-ordering neutrino masses, we determined the correlated $1 \sigma$ limits on the outer core composition and the diagonal and non-diagonal NSI coefficients.  In the non-diagonal case, the phases of the complex coefficients were marginalized.  The corresponding graphs are presented in Fig. \ref{fig:compo_eps}, where the regions to the right of the vertical dashed line correspond to the standard composition modified by the addition of hydrogen. The limits on the composition change due to modifications of the flavor probability transitions caused by NSI. This is especially notorious in the graphs of Fig. \ref{fig:epsNP_diag_c1_NO} where a positive (negative) correlation is observed between $\epsilon_{\tau\tau}-\epsilon_{\mu\mu}$ ($\epsilon_{ee}-\epsilon_{\mu\mu}$) and $(Z/A)_{oc}$, or even better, in the graphs of Figs. \eqref{fig:thirtyyearsexpositions_a} and \eqref{fig:thirtyyearsexpositions_b}, in terms of the hydrogen fraction. As we see, extreme values of $\digamma_{\!\!\chic H}$ are feasible in conjunction with large values of the diagonal coefficients. As an indication of the correctness of our results, we note that for $\epsilon_{\alpha \beta}$ equal to zero, we recover the outer core composition range obtained by considering only SM interactions \cite{DOlivoSaez:2022vdl}. Since our bounds on the NSI coefficients are of the same order of magnitude as those in the literature, one might infer that more restrictive bounds on these coefficients might be required to obtain reliable values of the hydrogen content in the Earth's outer core from an oscillation tomographic search.

\section*{Acknowledgments}
This work was supported by UNAM PASPA-DGAPA, CONAHCYT (Mexico) through grant CF-2023-I-1169,
and the Consejo Nacional de Investigaciones Cientificas y Tecnicas (CONICET) 
and the UNMDP Mar del Plata, Argentina.

\clearpage
\newpage
\mbox{}
\clearpage

\bibliographystyle{naturemag}
\bibliography{Nsi.bib}
\end{document}